\documentclass[aps,prb,reprint,floatfix,notitlepage, superscriptaddress]{revtex4-2}
\usepackage[utf8]{inputenc}
\usepackage{graphicx}
\usepackage{amsmath}
\usepackage{amssymb}
\usepackage{commath}
\usepackage{braket}
\usepackage{hyperref}
\usepackage{subfigure}
\usepackage[usenames,dvipsnames]{color}

\usepackage{booktabs} 
\usepackage{tabularx} 
\usepackage{array} 
\usepackage{longtable}

\usepackage[mathscr]{euscript}
\usepackage{bm}
\usepackage{dcolumn}
\usepackage{tikz,pgfplots}
\pgfplotsset{compat=1.18}
\usetikzlibrary{decorations.pathmorphing,patterns,positioning}
\usetikzlibrary{tikzmark}
\usetikzlibrary{shapes,shadows,arrows}

\usepackage{xcolor}
\definecolor{gatecolor}{HTML}{6FA4FF}

\newcommand\beq{\begin{equation}}
\newcommand\eeq{\end{equation}}
\newcommand\bea{\begin{eqnarray}}
\newcommand\eea{\end{eqnarray}}

\newcommand{\Tr}{{\rm Tr}}

\newcommand{\identity}{1\!\!1}

\usepackage{comment}
\usepackage[normalem]{ulem}

\newcommand{\relmiddle}[1]{\mathrel{}\middle#1\mathrel{}}

\begin{document}
	
\title{Local spreading of stabilizer R\'enyi entropy in a brickwork random Clifford circuit}

\author{ Somnath Maity}
\email{somnath.maity@riken.jp}
\affiliation{Nonequilibrium Quantum Statistical Mechanics RIKEN Hakubi Research Team, RIKEN Pioneering Research Institute (PRI), 2-1 Hirosawa, Wako, Saitama 351-0198, Japan}

\author{Ryusuke Hamazaki}
\affiliation{Nonequilibrium Quantum Statistical Mechanics RIKEN Hakubi Research Team, RIKEN Pioneering Research Institute (PRI), 2-1 Hirosawa, Wako, Saitama 351-0198, Japan} 
\affiliation{RIKEN Center for Interdisciplinary Theoretical and Mathematical Sciences (iTHEMS), 2-1 Hirosawa, Wako, Saitama 351-0198, Japan}

    
\date{\today}

\begin{abstract}
    Nonstabilizerness, or magic, constitutes a fundamental resource for quantum computation and a crucial ingredient for quantum advantage. Recent progress has substantially advanced the characterization of magic in many-body quantum systems, with stabilizer Rényi entropy (SRE) emerging as a computable and experimentally accessible measure. In this work, we investigate the spreading of SRE in terms of single-qubit reduced density matrices, where an initial product state that contains magic in a local region evolves under brickwork random Clifford circuits. For the case with Haar-random local Clifford gates, we find that the spreading profile exhibits a diffusive structure within a ballistic light cone when viewed through a normalized version of single-qubit SRE, despite the absence of explicit conserved charges. We further examine the robustness of this non-ballistic behavior of the normalized single-qubit SRE spreading by extending the analysis to a restricted Clifford circuit, where we unveil a superdiffusive spreading. Finally, we discuss that a similar non-ballistic spreading within the light cone is found for another indicator of the magic, i.e., the robustness of magic.
\end{abstract}

\maketitle

\section{Introduction}
In the quest for quantum advantage, considerable effort has been made to identify and characterize the resources that underlie quantum computational power, often within the resource-theoretic frameworks~\cite{chitambar2019quantum}. Entanglement is one such fundamental resource, central not only to quantum computation but also to our conceptual understanding of quantum many-body physics~\cite{osterloh2002scaling, amico2008entanglement,abanin2019colloquium,nahum2017quantum,skinner2019mipt}.  While entanglement is a key signature of quantumness, it is not, by itself, sufficient to achieve a quantum advantage. The class of stabilizer states is a prime example of states that are efficiently simulatable on classical computers even though they have a high degree of entanglement~\cite{gottesman1998theory,Aaronson2004improve}. To identify the essential ingredients of universal quantum advantage, one must therefore look at a quantity distinct from entanglement~\cite{bravyi2005universal,bravyi2016improved,campbell2017roads}.  The nonstabilizerness, often referred to as ``quantum magic,"  then turns out to be another  fundamental ingredient of the universal quantum advantages. It quantifies the amount of non-Clifford resources required to perform a quantum operation or to prepare a quantum state~\cite{veitch2014resource}. Given that the magic represents a dimension of quantumness different from entanglement, elucidating its physical consequences in quantum many-body systems is essential.  

\begin{figure}[t]
\centering
\includegraphics[width=0.4\textwidth,height=.31\textwidth]{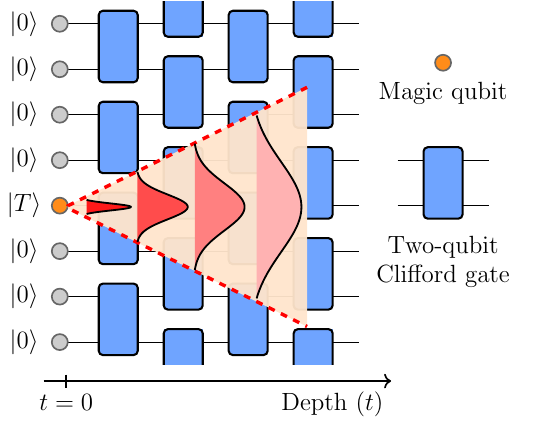}
\caption{An illustration of the results for local SRE spreading in a brickwork random Clifford circuit. A system of $L$ qubits at $t=0$ is prepared in an initial product state containing local magic (orange circle). The circuit comprises alternating layers of random two-qubit Clifford gates (blue square), uniformly chosen from the Clifford group $\mathcal{C}_2$, arranged in a staggered brickwork pattern. Under the time evolution, the spreading profile of the single-qubit averaged SRE exhibits a diffusive structure within the light cone. If we instead choose the two-qubit Clifford gates from restricted elements (see Fig.~\ref{fig:circuit_cnoths}) in the Clifford group, we find a superdiffusive structure.} 
\label{fig:rcbw}
\end{figure}

The properties of non-stabilizerness in quantum many-body physics had long been hard to address until recently, when the stabilizer Renyi entropy (SRE) was introduced as a computationally tractable measure~\cite{leone2022stabilizer}.  The SRE quantifies the spread of a quantum state in the Pauli-string basis via the Rényi entropy of its associated probability distribution. Notably, it is particularly efficient for many-body systems, as it avoids the costly optimization procedures required by the traditional magic monotones~\cite{bravyi2016trading,howard2017application,bravyi2019simulation, liu2022many}. Subsequent advancements in tensor network methods~\cite{haug2023quantifying,lami2023nonstabilizerness,lami2024unveiling,tarabunga2024nonstabilizerness_mps} and Monte-Carlo techniques~\cite{tarabunga2023many} have provided an effective toolkit for characterizing the nonstabilizerness of many-body ground states~\cite{oliviero2022magic,frau2024nonstabilizerness}. Several theoretical studies have recently addressed the quantum magic structure of various many-body systems, including spin chains~\cite{tarabunga2024critical,odavic2024stabilizer,turkeshi2025pauli,smith2025nonstabilizerness}, fermionic Gaussian states~\cite{collura2024quantum,tirrito2025magicphasetransitions}, field theories~\cite{white2021conformal,cao2024gravitational},  lattice gauge theories~\cite{tarabunga2023many}, and random circuits~\cite{zhang2024quantum,niroula2024phase,turkeshi2025magic,tirrito2024anticoncentration,haug2025probing,szombathy2024spectral,szombathy2025independent,tarabunga2024magic,bejan2024dynamical}. 

Despite these advances, the spreading behavior of the SRE remains to be fully understood.
The dynamical growth of SRE for the global pure state in Haar random quantum circuits has been studied in Ref.~\cite{turkeshi2025magic}, with emphasis on the exponential relaxation toward the saturation value. The equilibration occurs on timescales that scale logarithmically with the system size, reminiscent of anti-concentration and Hilbert space delocalization~\cite{dalzell2022random}. A universal behavior of magic growth, with a direct connection to transport properties, has been reported for U(1)-symmetric Hamiltonian dynamics in the XXZ spin chain~\cite{tirrito2025universalspreading}. However, these works focus only on the SRE of the global pure state, while the manner in which local SRE propagates through the system, along with the concurrent growth of entanglement, is still largely unexplored. 
In particular, even when we consider Clifford circuits, where the magic of the global state is unchanged, the distribution of SRE in local subsystems will show nontrivial dynamics.

In this work, we study the spatial spreading dynamics of the SRE for local subsystems under Clifford-circuit evolutions, starting from an initially prepared product state containing localized magic. To introduce the local magic, we apply a few single-qubit non-Clifford $T$ gates, ensuring that the initial state remains a superposition of only a small number of stabilizer states. The subsequent time evolution is generated by a brickwork random Clifford circuit, where the two-qubit gates are sampled uniformly from the full Clifford group (see Fig.~\ref{fig:rcbw}). Unlike the setups considered in many previous works, this simple setup enables us to efficiently simulate the nontrivial dynamics of magic even for large system sizes despite the rapid growth of entanglement.

Although the Clifford unitaries preserve the total amount of SRE in the system, the SRE for the reduced density matrix of the individual qubits, which we call the single-qubit SRE, nontrivially depends on time. Owing to the short-range nature of the brickwork quantum circuit, we find that SRE always spreads within a well-defined causal light cone. The single-qubit averaged SRE decays exponentially in time, reflecting the exponential growth of the single-qubit Pauli operators in operator space. 

Furthermore, by introducing a normalized version of the single-qubit averaged SRE,  we observe a nontrivial diffusive structure of its spreading, despite the absence of explicit conserved quantities.
That is, we numerically show that the spreading of the normalized single-qubit averaged SRE obeys the discrete diffusion equation, leading
to its diffusive profile within the light cone. 
Moreover, we examine the robustness of this non-ballistic structure of the single-qubit SRE spreading for a restricted Clifford circuit.
In this case, the diffusion equation does not hold; however, we still observe a non-ballistic behavior, represented by superdiffusive spreading.    

We also extend our analysis to a faithful magic monotone for mixed states, namely the log-free robustness of magic magic (LROM). We find that single-qubit LROM exhibits qualitatively similar non-ballistic spreading accompanied by the exponential decay. 

Moreover, in our main setup, the single-qubit SRE and LROM admit a natural operational interpretation: they quantify the probability of recovering the initially injected magic state at a given site in a specific realization of the random Clifford circuit. The single-qubit SRE provides an upper bound to this probability, and the single-qubit ROM is proportional to the probability.

The remainder of the paper is organized as follows. In Sec.~\ref {sec:pre}, we establish notations and introduce the SRE.  Section~\ref {sec:setup} describes the quantum circuit and the setup used to study the spreading of SRE. We then proceed to present the main results of the spreading of SRE in Sec.~\ref {sec:results}. Qualitative explanations based on the spreading of local Pauli operators and the analysis of a single Clifford gate are presented in Sec.~\ref {sec:pauli_spreading} and Sec.~\ref {sec:single_clifford}, respectively. In Sec.~\ref {sec:restricted_clifford}, we demonstrate the generality of the existence of the non-ballistic structure, extending our study to a restricted Clifford circuit. In Sec.~\ref{sec:rom}, we present results for the robustness of magic. A discussion of operational implications of our setup and the associated spreading dynamics is presented in Sec.~\ref{sec:implications}. Finally, we conclude in Sec.~\ref{sec:discussion} with a summary of key results and the possible future directions. Additional details, including the calculations of the SRE for a single magic state, elements in the Pauli spectrum, and the possible two-qubit Pauli spectra, are provided in Appendices ~\ref{app:magic_tstat}, ~\ref{app:ps}, and ~\ref{app:allowed_ps}, respectively.

\section{Preliminaries}\label{sec:pre}
Let us start by considering a system of $L$ qubits with a Hilbert space $\mathcal{H}= (\mathbb{C}^2)^{\otimes L}$ of dimension $d=2^L$. The $L$-qubit Pauli group comprises all possible Pauli strings with overall phases $i^{\phi}$, defined as
\begin{eqnarray}
     \Tilde{\mathcal{P}}_L= \left\{ i^{\phi} \sigma_{0}\otimes\sigma_{1}\otimes\cdots\otimes\sigma_{L-1}\right\},
\end{eqnarray}
where $\sigma_j = \{\identity_j, X_j, Y_j, Z_j\}$ are the Pauli matrices for site $j$ and $\phi=\{0,1,2,3\}$. Pauli strings without the global phases, $\mathcal{P}_L = \Tilde{\mathcal{P}}_L \backslash U(1)$, form an orthonormal basis for the space of linear operators on $\mathcal{H}$. 

The Clifford group is another important group that allows us to define the resource theory of quantum magic. It is defined as the normalizer of the $L$-qubit Pauli group:
\begin{eqnarray}\label{eq:cliff_op}
    \mathcal{C}_L := \left\{ C \in U(d) ~|~ CPC^\dagger  \in \Tilde{\mathcal{P}}_L, \forall P \in \Tilde{\mathcal{P}}_L \right\}.
\end{eqnarray}
The Clifford group is generated by $S = \sqrt{Z}$, $H = (X + Z)/\sqrt{2}$, and $CNOT$
gates and forms a unitary $3$-design for qubits, approximating Haar randomness up to third moments.

Given the elements of the Clifford groups, the set of stabilizer states on $\mathcal{H}$ is defined as the orbit of the Clifford group starting from the computational basis $\ket{0}^{\otimes L}$:
\begin{eqnarray}\label{eq:stab}
    STAB := \left\{ C \ket{0}^{\otimes L} : C \in \mathcal{C}_L \right\}.
\end{eqnarray}
Equivalently, a pure state $\ket{\psi}$ is a stabilizer state if there exists a subgroup, $S \subset \mathcal{P}_L$, of $d$ mutually commuting Pauli strings such that $\bra{\psi}P\ket{\psi}=\pm1$ for all $ P \in S$ and $\bra{\psi}P\ket{\psi}=0$ for $P \notin S$.

Despite their intricate architecture and the extensive entanglement generated by the Clifford operations, the stabilizer states alone are insufficient for achieving universal quantum computation. Since the set of stabilizer states is closed under Clifford operations, universality requires the inclusion of additional non-Clifford resources—either non-Clifford unitaries or non-stabilizer states. The resource that measures the amount of deviation from the stabilizer manifold is known as non-stabilizerness or quantum magic. A canonical example of a resource state enabling universality is the T-state, defined as 
\begin{eqnarray}\label{eq:tstate}
    \ket{T} = \frac{1}{\sqrt{2}}\left(\ket{0} + e^{i\pi/4} \ket{1}\right),
\end{eqnarray}
which can be prepared by applying a Hadamard gate followed by the non-Clifford $T=\mathrm{diag}(1, e^{i\pi/4})$ gate to the stabilizer state $\ket{0}$.

\subsection{Stabilizer R\'enyi entropy}
To quantify the amount of non-stabilizerness in a scalable manner, we mainly focus on the stabilizer R\'enyi entropy (SRE), first introduced in Ref.~\cite{leone2022stabilizer}.  For a pure state $\ket{\psi}$, the square normalized expectation value of the Pauli strings $P\in \mathcal{P}_L$,
\begin{equation}\label{eq:prob_dis}
    \Theta_{\psi} \left(P\right)=\frac{1}{d}\lvert \bra{\psi}P\ket{\psi}\rvert^2,
\end{equation}
forms a well-defined probability distribution since $\Theta_{\psi} \left(P\right)\geq0$ and $ \sum_{P \in \mathcal{P}_L}\Theta_{\psi} \left(P\right)=1$.  With the probability distribution $\Theta_{\psi} \left(P\right)$, the SRE of order $\alpha$ for pure states is defined as 
\begin{eqnarray}\label{eq:sre_alpha}
M^{(\alpha)}(\ket{\psi})&=& \frac{1}{1-\alpha}\log \left(\frac{1}{d}\sum_{P \in \mathcal{P}_L}\lvert \bra{\psi}P\ket{\psi}\rvert^{2\alpha}\right),
\end{eqnarray}
which corresponds to the $\alpha$-R\'enyi entropies of the probability distribution shifted by $-\log(d)$. From a resource-theoretic perspective, the SRE serves as a good magic monotone for $\alpha\ge2$~~\cite{leone2022stabilizer,haug2023stabilizer,leone2024stabilizer}. In particular, the SRE satisfies the following key properties:
(i) faithfulness: $M^{(\alpha)}(\ket{\psi})=0$ if and only if $\ket{\psi}$ is a stabilizer state, (ii) invariance under free (Clifford) operations: $M^{(\alpha)}(\Gamma\ket{\psi})=M^{(\alpha)}(\ket{\psi})$ for all $\Gamma \in \mathcal{C}_L$, and (iii)
additivity: $M^{(\alpha)}(\ket{\psi}\otimes\ket{\phi})=M^{(\alpha)}(\ket{\psi})+M^{(\alpha)}(\ket{\phi})$.

\subsection{SRE for mixed states}
The definition of SRE can also be extended to mixed states~\cite{leone2022stabilizer}, including reduced density matrices of subsystems in many-body systems. In this case, the state-dependent normalized probability distribution is defined as 
\begin{eqnarray}
    \Theta_{\rho}(P)=\frac{1}{2^{n}} \frac{\Tr(\rho P)^2}{\Tr(\rho ^2)},
\end{eqnarray}
for an $n$-qubit density matrix $\rho$. This quantity is non-negative and normalized, ensuring that it can be treated analogously to the pure-state distribution in Eq.~\eqref{eq:prob_dis}. The corresponding SRE of order $\alpha$ for mixed states is then given by~\cite{tarabunga2025efficient},
\begin{eqnarray}\label{eq:sre_mixed}
    M^{(\alpha)}(\rho) = \frac{1}{1-\alpha}\left(\log A_{\alpha}(\rho) + S_2(\rho)\right),
\end{eqnarray}
where $S_2(\rho)=-\log(\Tr[\rho^2])$ denotes the second R\'enyi entropy, and $A_{\alpha}(\rho)$ represents the $\alpha$-moment of the Pauli spectrum, defined as 
\begin{eqnarray}
    A_{\alpha}(\rho) = \frac{1}{2^n} \sum_{P\in \mathcal{P}_n} \lvert \Tr[\rho P] \rvert^{2\alpha}.
\end{eqnarray}In this work, we focus exclusively on the case $\alpha=2$. The second-order stabilizer R\'enyi entropy ($2$-SRE) for the density matrix $\rho$ takes the following compact form,
\begin{eqnarray}\label{eq:sreden}
    M^{(2)}(\rho) = -\log \left( \frac{\sum_{P \in \mathcal{P}_n} \lvert c_P \rvert^4}{\sum_{P \in \mathcal{P}_n} \lvert c_P \rvert^2} \right),
\end{eqnarray}
where $c_P = \Tr[\rho P]$ are the Pauli expansion coefficients of the state $\rho$. The above quantity also satisfies the key properties of the measure for non-stabilizerness as discussed above for the pure state~\footnote{More precisely, the faithful property holds true when one defines that the mixed stabilizer state is given in the form $\rho=\identity/2^n+1/2^n\sum_{P\in G}\phi_PP$, where $G$ is a subset of  $\mathcal{P}_n$ satisfying $1<|G|<2^n-1$ and $\phi_P=\pm 1$~\cite{leone2022stabilizer}. }, though its strict monotonicity under resource operations remains an open theoretical issue~\cite{haug2023stabilizer,leone2024stabilizer}. Nonetheless, the $2$-SRE, $M^{(2)}(\rho)$, is a commonly used scalable measure of magic for mixed states~\cite{oliviero2022magic,tarabunga2023many,tarabunga2024critical,frau2024nonstabilizerness,lopez2024exact,frau2025stabilizer}.
In the following, we simply write $M^{(2)}$ as $M$.

\section{Setup}\label{sec:setup}
We now explain our setup.
Given an array of $L$ qubits, we build quantum circuits with a brickwork structure of odd and even layers of local (two-qubit) Clifford gates $C \in \mathcal{C}_2$. Each gate is
chosen randomly with uniform probability from all possible two-qubit Clifford gates from the Clifford group $\mathcal{C}_2$ ($|\mathcal{C}_2| = 11520$)~\cite{Bravyi2021}. The unitary time evolution operator for the random brickwork Clifford circuit reads as
\begin{equation}\label{eq:rcbw}
    \mathcal{U}(t) = U(t,t-1)\cdots  U(2,1)U(1,0),
\end{equation}
where $U(t',t'-1)$ is the evolution operator for the $t'$-th layer (time step) in the circuit. Numbering the qubits as $i=0, 1 ,2,  \dots L-1$, the unitary operator $U(t',t'-1)$ is given by
\begin{equation}\label{eq:brickwork}
    U(t',t'-1) = \begin{cases} \bigotimes_{k=0}^{L/2 -1} C_{2k,2k+1}^{(t')} ~~~~~\text{if $t'$ is odd},\\ 
    \bigotimes_{k=0}^{L/2 -1} C_{2k+1,2k+2}^{(t')} ~~\text{if $t'$ is even},
    \end{cases}
\end{equation}
where $C_{i,j}^{(t')}$ is a two-qubit Clifford gate that acts on sites $i$ and $j$ at time $t'$ and is chosen randomly from a uniform distribution on a two-qubit Clifford group $\mathcal{C}_2$. In our numerical calculations, the random two-qubit Clifford gates are sampled using the Qiskit library~\cite{qiskit2024} based on the method in Ref.~\cite{Bravyi2021}. We consider the periodic boundary condition $C_{L-1,L}^{(t')} = C_{0,L-1}^{(t')}$. The brickwork Clifford circuit is graphically illustrated in Fig.~\ref{fig:rcbw}.

At $t=0$, the system is prepared in the following low-magic initial state 
\begin{eqnarray}\label{eq:ini_state}
\ket{\psi_0}&=&\ket{0}\otimes\cdots\ket{0}\otimes\ket{T}\otimes\ket{0}\cdots\otimes\ket{0},
\end{eqnarray}
where the local magic is injected at $i=m$ through a single-qubit $\ket{T}$ state embedded within a product state of $\ket{0}$ states for the remaining qubits. For brevity, we denote this product state as $\ket{00\cdots 0T0 \cdots 0}$. In a quantum circuit, such a state is prepared by applying a Hadamard gate $H$ followed by a $T$ gate on the qubit at site $m$ in the stabilizer state $\ket{0}^{\otimes L}$. The initial state can equivalently be expressed as a superposition of two stabilizer states: 
\begin{equation}\label{eq:ini_state_super}
    \ket{\psi_0} = \frac{1}{\sqrt{2}} \ket{0\cdots 000 \cdots 0} + \frac{e^{i\pi/4}}{\sqrt{2}} \ket{0\cdots 010 \cdots 0}.
\end{equation}

The system then evolves under the brickwork random Clifford circuit, such that the state at time $t$ is given by $\ket{\psi_t}=\mathcal{U}(t)\ket{\psi_0}$. We employ an efficient algorithm to numerically evaluate the SRE by representing Pauli strings in the symplectic (binary vectors) form and updating them efficiently under the local Clifford operations~\cite{Aaronson2004improve}, without invoking any approximations.  To this end, we work in the Heisenberg picture, where for a given Pauli string operator $P$, we compute its time evolution $P(t)=\mathcal{U}(t)^{\dagger}P\mathcal{U}(t)$, and subsequently evaluate its expectation value on the initial state $\ket{\psi_0}$ to calculate the time-dependent expectation value $\langle P(t)\rangle=\bra{\psi_0}\mathcal{U}(t)^{\dagger}P\mathcal{U}(t)\ket{\psi_0}$. Since Clifford unitaries map a Pauli string onto another Pauli string, the expectation value of the time-evolved Pauli string $P(t)$ can be easily calculated on the low-magic initial state defined in Eq.~\eqref{eq:ini_state_super}. Consequently, the SRE defined in Eq.~\eqref{eq:sreden} can be calculated efficiently and exactly from the expectation values of the time-evolved Pauli strings, as discussed in the next section for the single-qubit case.

\begin{figure}[]
    \centering
    \includegraphics[width=0.45\textwidth,height=.35\textwidth]{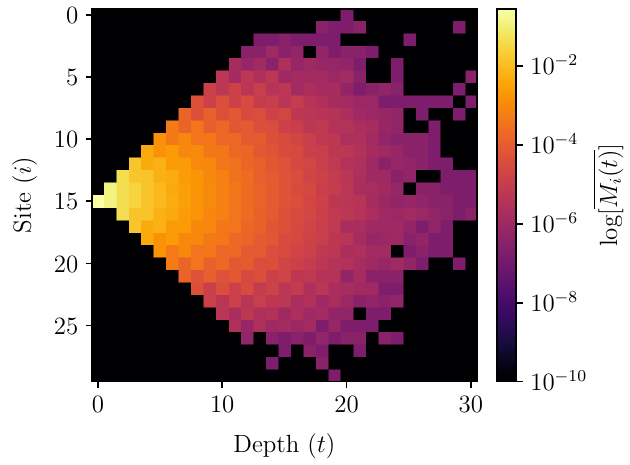}
    \caption{Spreading of the averaged single-qubit SRE, $\overline{M_i(t)}$, as a function of site $i$ and depth (time) $t$ of the circuit. The initial state $\ket{\psi_0}$ contains local magic at site $i=15$ of a system of total size $L=30$. As the quantity $\overline{M_i(t)}$ is exponentially decaying with time, we plot its logarithmic value for better visualization. The average is taken over $10^6$ circuit realizations.}
    \label{fig:sre_spread}
\end{figure}

\section{Numerical Results}\label{sec:results}

\subsection{Framework}
To analyze the spreading of SRE, we evaluate the $2$-SRE of the reduced density matrix of each qubit at every time step, which we refer to as the single-qubit SRE.  
For a given qubit at site $i$, we introduce the corresponding local Pauli strings, 
\begin{eqnarray}
    P_{\sigma_i}=\left\{\identity_0\identity_1\cdots\identity_{i-1}\sigma_{i}\identity_{i+1}\cdots\identity_{L-1}\right\},
\end{eqnarray}
where $\sigma_{i} = \{\identity_i, X_i, Y_i, Z_i\}$. General Clifford circuits exhibit a particularly simple structure for their operator spreadings:  Pauli operators evolve within the Pauli group, with mutual commutation relations preserved. Thus, the local Pauli string $P_{\sigma_i}$ after time steps $t$ is  given by another Pauli string,
\begin{eqnarray}
    P_{\sigma_i}(t) = \mathcal{U}(t)^{\dagger}P_{\sigma_i}\mathcal{U}(t) \in \mathcal{P}_L.
\end{eqnarray}
We then calculate the expectation values of all the time-evolved Pauli strings on the initial state $\Tr[\rho_i(t)P_{\sigma_i}]= c_{\sigma_i}(t)=\bra{\psi_0} P_{\sigma_i}(t) \ket{\psi_0}$. This yields the time-dependent Pauli spectrum $\left\{\lvert c_{\sigma_i}(t)\rvert\right\}$ corresponding to the reduced density matrix $\rho_i(t)$ of $\ket{\psi_t}$ for each qubit $i$~\cite{lopez2024exact}. It can then be expressed as
\begin{equation}
    \rho_{i} (t) = \frac{1}{2} \sum_{\sigma_i \in \{\identity, X, Y, Z\}} c_{\sigma_i}(t) P_{\sigma_i}.
\end{equation}
Using these Pauli expansion coefficients in Eq.~\eqref{eq:sreden}, we can calculate the time-dependent single-qubit SRE for site $i$ as 
\begin{equation}\label{eq:sre_coeff}
    M_i(t) = - \log \left( \frac{\sum_{\sigma_i \in \{\identity, X, Y, Z\}} \lvert c_{\sigma_i}(t)\rvert^{4}}{\sum_{\sigma_i \in \{\identity, X, Y, Z\}} \lvert c_{\sigma_i}(t)\rvert^{2}} \right).
\end{equation}
Finally, we calculate the averaged single-qubit SRE, $\overline{M_i(t)}=\mathbb{E}_{\mathrm{Cl}} [M_i(t)]$, over realizations of the random Clifford circuit.

\begin{figure*}[]
    \centering
    \includegraphics[width=.85\textwidth,height=.29\textwidth]{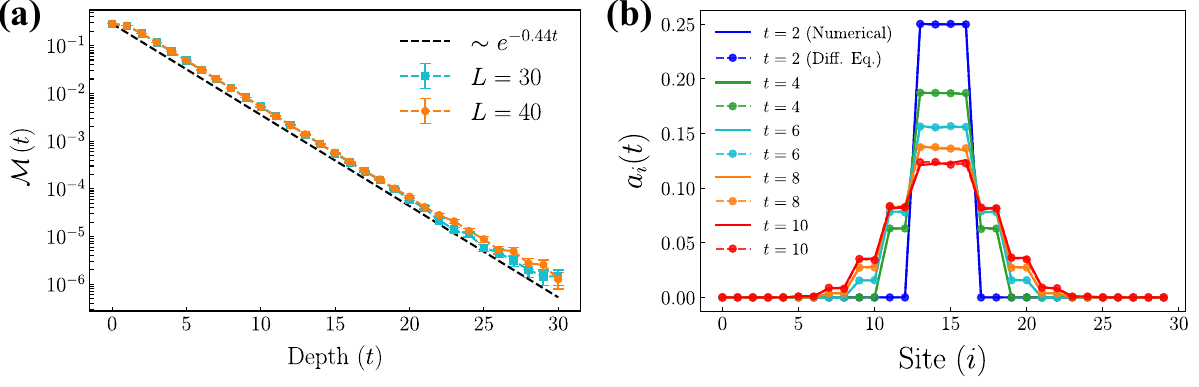}
    \caption{(a) Exponential decay of the sum of averaged single-qubit SREs, $\mathcal{M}(t)$, as given in Eq.~\eqref{eq:expo_decay}. The decay rate $\Gamma=0.44$ is indicated by the black dashed reference line with slope $-0.44$. (b) Variation of the normalized single-qubit SREs, $a_i(t)$, as a function of site $i$ at different time steps $t=2, 4, 6, 8, 10$. The solid lines represent numerical data, while the dashed lines show the corresponding values obtained from the discrete diffusion equation (Eq.~\eqref{eq:dis_diff}) using the numerical data from the previous time step. The description by the diffusion equation almost agrees with the actual dynamics.  The total system size is $L = 30$. In both panels, all results are averaged over $10^6$ realizations.}
    \label{fig:diff_spread_rcbw}
\end{figure*}

\subsection{Numerical Results}

We begin by presenting numerical results for the spatial and temporal spreading of $2$-SRE under the brickwork Clifford circuit. In Fig.~\ref{fig:sre_spread}, we plot the space-time spreading of the averaged single-site SRE, $\overline{M_i(t)}$, as a function of site index $i$ and time $t$.

At $t=0$, the system is prepared in a product state, where all qubits are in stabilizer states except for a single qubit $i=m$ initialized at $T$-state. Consequently, only this site exhibits a nonzero single-qubit SRE, with a value of $\log(4/3)$ (see Appendix~\ref{app:magic_tstat}). All the other sites have zero SRE, and the total SRE for the global state is simply the sum of the SREs of individual sites and is equal to $\log(4/3)$.

As the system evolves under a locally interacting brickwork random Clifford circuit, correlations begin to build up and propagate through the system. Due to the locality of the circuit, the propagation of correlations is constrained within a light cone. The spreading of SRE in the circuit is indeed restricted within an outward light cone
from the initial magic qubit, as shown in Fig.~\ref{fig:sre_spread}. While the global state remains pure and undergoes a unitary Clifford evolution, which preserves the total amount of SRE, the local structure of SRE changes across the system. 
Specifically, averaged single-qubit SRE $\overline{M_i(t)}$ rapidly decreases with time once the light cone reaches the site $i$ (i.e., when $t \geq l$, where $l$ is the distance of the qubit $i$ from the qubit $m$, where nonzero magic is initially inserted). Note that the SRE  at long times exhibits very small values, which lead to non-negligible statistical fluctuations for the finite number of samples, as observed in Fig.~\ref{fig:sre_spread}.

Although the total SRE of the global pure state remains conserved, the sum of the averaged single-qubit SRE, 
\begin{align}
\mathcal{M}(t) = \sum_{i = 0}^{L-1} \overline{M_i(t)}, 
\end{align}
decays exponentially in time after some transient time:
\begin{equation}\label{eq:expo_decay}
    \mathcal{M}(t) \sim \exp (-\Gamma t),
\end{equation}
with $\Gamma = 0.44$ being the decay rate extracted numerically, as shown in Fig.~\ref{fig:diff_spread_rcbw}(a). We emphasize that the above equation is valid only before the light cone covers the whole system, as the ballistic spread of the Pauli operator stops and starts overlapping for the periodic boundary condition. 
After the time when the light cone covers the whole system, deviation from Eq.~\eqref{eq:expo_decay} starts to increase,
which eventually becomes visible for long times, as shown in Fig.~\ref{fig:diff_spread_rcbw}(a). 

To further probe the nature of this spreading, we introduce a normalized version of the averaged single-qubit SRE,
\begin{equation}
    a_i(t)= \frac{\overline{M_i(t)}}{\mathcal{M}(t)},
\end{equation}
which characterizes how the local magic is spatially distributed at a given time $t$.
Remarkably, we numerically find that the normalized averaged SRE obeys the recurrence relation of a discrete random walk,
\begin{eqnarray}\label{eq:dis_diff}
    a_i(t) = \frac{1}{2} \left[ a_{i-1}(t-1) + a_{i+1}(t-1) \right],
\end{eqnarray}
for the sites within the light cone. This is shown in Fig.~\ref{fig:diff_spread_rcbw}(b). In the long-time and large-scale (hydrodynamic) limit, the above equation converges to the diffusion equation
\begin{equation}
    \frac{\partial a(x,t)}{\partial t} = D \frac{\partial^2 a(x,t)}{\partial x^2},
\end{equation}
with the diffusion constant $D=1/2$. Therefore, the spatial distribution of the normalized SRE spreads diffusively as the width of the distribution broadens in time as $\sigma(t) \sim\sqrt{t}$ around the mean position at the location of the magic qubit.

The above results demonstrate that, once normalized, the SRE reveals a diffusive structure on top of the light cone spreading. This reflects an emergent hydrodynamic-like spreading of SRE, even though the underlying dynamics are purely unitary and generated by Clifford operations without conserved quantities. 
We emphasize that this behavior is distinct from the diffusive broadening in the operator spreading  in  Haar-random circuits without conserved quantities~\cite{Keyserlingk2018,nahum2018operator}.
In fact, these results discuss the diffusive spreading of the operator front near the light cone, while 
the diffusive spreading of the normalized single-qubit SRE found here occurs at the bulk well inside the light cone.

\section{Single-qubit Pauli spectrum via Pauli-operator spreading}\label{sec:pauli_spreading}
To gain insight into the numerical results of SRE spreading, we first examine the operator spreading of the single-site Pauli operators $\sigma_i$ under the Clifford circuit and, consequently, their expectation values in the initial state, which together constitute the Pauli spectrum $\{\lvert c_{\sigma_i}(t)\rvert\}$. 

Under the action of a Clifford unitary, any Pauli string is mapped to another
Pauli string. In a locally interacting brickwork Clifford circuit, the time evolution of an initially local Pauli operator typically results in a nonlocal Pauli string due to the linear spreading of operators. The time-evolved Pauli string $P_{\sigma_i}(t) = \mathcal{U}(t)^{\dagger}P_{\sigma_i}\mathcal{U}(t)$ has a nontrivial support exactly on the backward light cone of the spacetime:
\begin{eqnarray}
    P_{\sigma_i}(t) = \bigotimes_{\mu=0}^{L-1}\sigma_\mu,
\end{eqnarray}
with 
\begin{align*}
    \sigma_\mu &= \{ \identity_\mu, X_\mu, Y_\mu, Z_\mu \}, &~\text{for}~ \mu \in [s(i,t),\ell(i,t)], \\
    \sigma_\mu &=  \identity_\mu, &~\text{for}~ \mu \notin [s(i,t),\ell(i,t)].
\end{align*}
The indices $s(i,t)$ and $\ell(i,t)$ denote the smallest and largest sites, respectively, for the interval that spans the backward light cone of the site $i$ located at depth $t$. Owing to the brickwork structure of the circuit, these boundaries depend on the parities of both $i$ and $t$. For odd $i$, the backward light spans the interval $[s(i,t),\ell(i,t)]=[i-t, i+t-1]$ when $t$ is odd, and $[i-t+1, i+t]$ when $t$ is even. Conversely, for even $i$, the interval is $[i-t+1, i+t]$ when $t$ is odd, and $[i-t, i+t-1]$ when $t$ is even.
For clarity, we omit the identity operators for the sites outside the light cone and write the time-evolved operator using only the nontrivial segment of the Pauli string,
\begin{eqnarray}
    P_{\sigma_i}(t) = \sigma_{s(i,t)}\cdots \sigma_{i-1}\sigma_i \sigma_{i+1}\cdots \sigma_{\ell(i,t)}.
\end{eqnarray}
Thus, the spatial extent of the time-evolved Pauli strings typically grows linearly in time, reaching $2t$ after $t$ time steps, while the total number of distinct Pauli strings increases exponentially as $4^{2t}$. 

If the $T$-state in $\ket{\psi_0}$ is located at site $m$ that lies within the backward light cone of the site $i$, i.e, $m \in [s(i,t),\ell(i,t)]$,
the expectation values of all possible Pauli strings have the following structure. The nonzero expectation values of the time-evolved Pauli strings take the form
\begin{eqnarray}
    \langle A_{s(i,t)}\cdots A_{m-1}\identity_m A_{m+1}\cdots A_{\ell(i,t)} \rangle &=& 1, \\
    \langle A_{s(i,t)}\cdots A_{m-1}B_m A_{m+1}\cdots A_{\ell(i,t)} \rangle &=& \pm\frac{1}{\sqrt{2}},
\end{eqnarray}
where $A_i = \{\identity, Z\}$, $B_i = \{X,Y\}$, and $\langle \cdot \rangle$ denotes the expectation value over the initial state $ \ket{\psi_0}$.  All the other Pauli strings yield vanishing expectation values (see Appendix~\ref{app:ps}). Conversely, if the site $m$ lies outside the light cone, the Pauli spectrum remains trivial, with only nonzero terms given by
\begin{eqnarray}
    \langle A_{s(i,t)}\cdots A_{i-1}A_i A_{i+1}\cdots A_{\ell(i,t)} \rangle &=& 1,
\end{eqnarray}
corresponding to a vanishing SRE for the qubit at site $i$. 

For a nontrivial Pauli spectrum, i.e., when $m\in [s(i,t),\ell(i,t)]$, the total number of nonzero elements among all $4^{2t}$ Pauli strings is thus given by $3\cdot 2^{2t-1}$. Consequently, the ratio for the number of elements with nonzero expectation value to the number of elements having zero expectation value decays exponentially with time as $\sim 2^{-2t}$. For a single realization of the circuit of depth $t$, the Pauli spectrum $\{\lvert c_{\sigma_i}(t)\rvert\}$ for a single qubit is determined by the expectation values of only four out of $4^{2t}$ possible Pauli strings. Hence, the probability of obtaining a nontrivial Pauli spectrum-- yeiding a nonzero SRE, such as $\{1, 1/\sqrt{2}, 1/\sqrt{2},0\}$ or $\{1, 1/\sqrt{2},0,0\}$ (see Appendix.~\ref{app:allowed_ps}])-- decays exponentially in time. At late times, the Pauli spectrum for each qubit is mostly given by $\{1,0,0,0\}$, corresponding to a zero SRE value. This behavior qualitatively accounts for the exponential decay of the total single-qubit averaged SRE $\mathcal{M}(t)$ in Eq.~\eqref{eq:expo_decay}.

\section{Two-qubit Pauli spectrum across a single Clifford gate}\label{sec:single_clifford}
\begin{figure}
\centering
\includegraphics[width=0.4\textwidth,height=.3\textwidth]{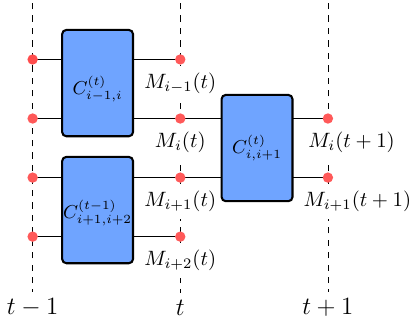}
\caption{Input (time $t$) and output (time $t+1$) single-qubit SREs across a Clifford gate $C_{i,i+1}^{(t)}$ acting on neighboring sites $i$ and $i+1$. The averaged output values  are equal $\overline{M_i(t+1)}=\overline{M_{i+1}(t+1)}$. Similarly, at time step $t$, the outputs of the gates in the previous layers satisfy $\overline{M_{i-1}(t)}=\overline{M_{i}(t)}$ and $\overline{M_{i+1}(t)}=\overline{M_{i+2}(t)}$.} 
\label{fig:single_clifford}
\end{figure}

Next, we examine the diffusive structure in the spreading of the normalized single-qubit SRE in more detail.
For this purpose, let us consider a single two-qubit Clifford gate $C_{i,i+1}^{(t)}$ acting on a pair of neighboring sites $i$ and $i+1$ from a layer $t$ of the brickwork circuit. In our setup of the brickwork circuit, the gate belongs to the odd (even) layer $t$, corresponding to an even (odd) pair of sites, respectively (see Fig.~\ref{fig:single_clifford}). 
There are sixteen Pauli strings acting on this two-site Hilbert space, and their time-evolved expectation values are given by,
\begin{eqnarray}
    c_{\sigma_i, \sigma_{i+1}} (t)= \bra{\psi_0}P_{\sigma_i, \sigma_{i+1}}(t)\ket{\psi_0},
\end{eqnarray}
where $P_{\sigma_i, \sigma_{i+1}}(t)=\mathcal{U}(t)^{\dagger}P_{\sigma_i, \sigma_{i+1}}\mathcal{U}(t)$ with $P_{\sigma_i, \sigma_{i+1}}$ indicating the local two-qubit Pauli strings. This will give the two-qubit Pauli spectrum $\{\lvert c_{\sigma_i \sigma_{i+1}}(t)\rvert\}$. 

\begin{figure}[]
    \centering
    \includegraphics[width=0.4\textwidth,height=.35\textwidth]{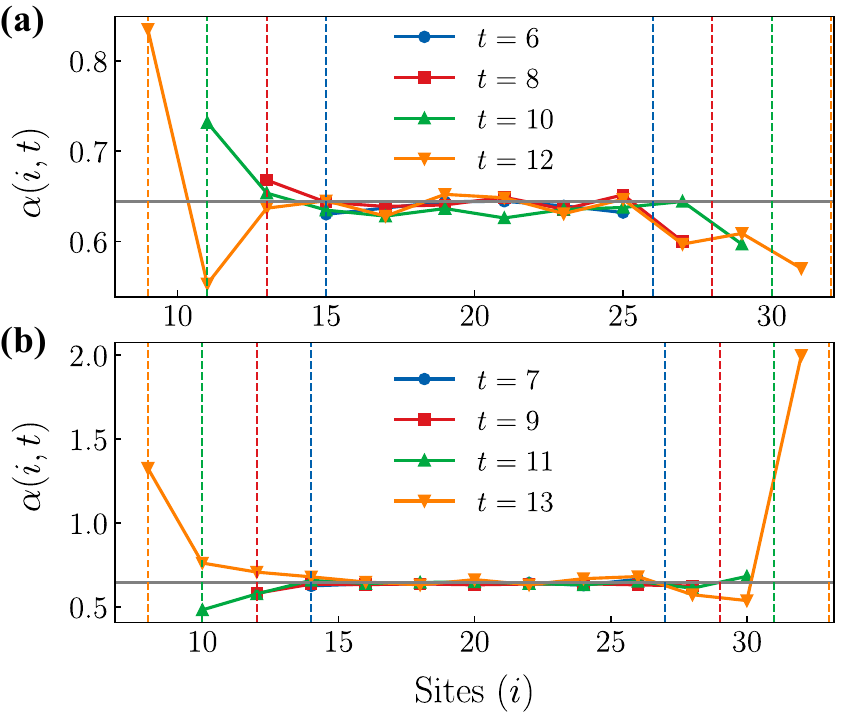}
    \caption{Variation of $\alpha(i,t)$ as a function of the Clifford gate position $i$ within the lightcone at different (a) even time steps $t = 6,8,10,12$, and (b) odd time steps $t = 7, 9, 11, 13$. In both panels, the position of the light cone boundaries corresponding to each time step is shown by the vertical dashed lines of the same color. The results show that the quantity $\alpha(i,t)$ is approximately independent of both $i$ and $t$, and equal to $e^{-\Gamma}=0.644$ (indicated by the solid gray line), at least well inside the light cone. }
    \label{fig:ratio_inout}
\end{figure}

Under the two-site Clifford operation, the number and values of the nonzero elements in the two-qubit Pauli spectrum remain unchanged; they are merely redistributed so that the SRE of the two-qubit reduced density matrix is conserved at the successive time steps $t$ to $t+1$. However, the redistribution of the two-qubit Pauli spectrum can change the single-qubit SRE of the individual qubit at the output, since the single-qubit SREs for the qubits at sites $i$ and $i+1$ depend on the coefficients, 
\begin{eqnarray}
    \left\{c_{\identity\identity}(t+1), c_{X\identity}(t+1), c_{Y\identity}(t+1), c_{Z\identity}(t+1)\right\},
\end{eqnarray}
and
\begin{eqnarray}
    \left\{c_{\identity\identity}(t+1), c_{\identity X}(t+1), c_{\identity Y}(t+1), c_{\identity Z}(t+1)\right\},
\end{eqnarray}
respectively.
Since the group of two-qubit Clifford gates is invariant under multiplication by the two-qubit swap gate, which is itself a Clifford gate, the single-qubit SREs of the two qubits at the output are, on average, equal:
\begin{eqnarray}\label{eq:equal_output}
     \overline{ M_i(t+1)} = \overline{ M_{i+1} (t+1) }.
\end{eqnarray}
Note that the above equality is only true at the output of each Clifford gate, but the averaged single-qubit SREs at the two inputs of the Clifford gate are typically different, as they are the output of two different neighboring gates.

\begin{figure*}[]
    \centering
    \includegraphics[width=.83\textwidth,height=.3\textwidth]{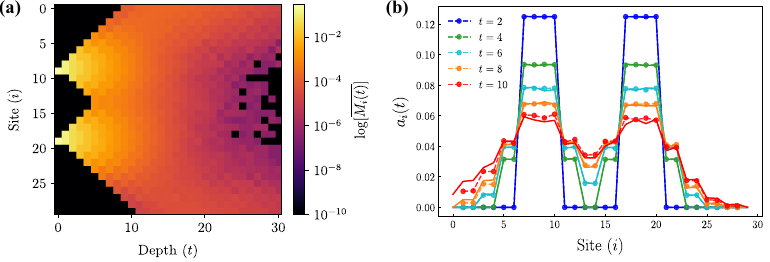}
    \caption{(a) Spreading of the averaged single-qubit SRE, $\overline{M_i(t)}$, and (b) diffusive behavior of the normalized single-qubit SRE, $a_i(t)$, for an initial state containing two magic ($T$) states located at the sites $i=9$ and $i=19$. The results are averaged over $10^6$ realizations, and the total system size is $L=30$. In (b), the solid lines represent the numerical data, while the dashed lines with markers show the corresponding values obtained from the discrete diffusion equation using data from the previous time step. This agreement confirms the diffusive structure in the SRE spreading.}
    \label{fig:sre_spread_tq}
\end{figure*}

With the above condition, the diffusive structure observed in the spreading of the normalized single-qubit averaged SRE can be understood as follows. We consider the ratio between the sum of averaged single-qubit SREs at the output and that at the input of the Clifford gate $C_{i,i+1}^{(t)}$ as,
\begin{eqnarray}\label{eq:single_gate_decay}
     \frac{\overline{ M_i(t+1)} + \overline{ M_{i+1} (t+1) }}{\overline{ M_i(t)} + \overline{ M_{i+1} (t) }} =  \alpha(i,t).
\end{eqnarray}
In general, the quantity $\alpha(i,t)$ depends on both the position $i$ and time $t$. We assume that $\alpha(i,t)$ is always finite, such that the sum of the SREs at the output is zero whenever the Clifford gate is located outside the light cone, where the sum of the input SREs is zero. Note that the sum in the above equation is over only two sites on which the single Clifford gate is located, and it is different from Eq.~\eqref{eq:expo_decay}, which concerns the sum over all qubits. For two successive time steps $t$ and $t+1$, the Eq.~\eqref{eq:expo_decay} takes the following form:
\begin{equation}\label{eq:totalsre_ratio}
    \frac{\mathcal{M}(t+1)}{\mathcal{M}(t)}=e^{-\Gamma}.
\end{equation}

Nontrivially, we numerically find that the coefficients $\alpha(i,t)$ are approximately independent of $i$ and $t$, and it is equal to $e^{-\Gamma}$. 
We illustrate this fact in Fig.~\ref{fig:ratio_inout}. We plot the ratio $\alpha(i,t)$ 
for all the Clifford gates located within the light cone at different time steps. Note that for even time steps [Fig.~\ref{fig:ratio_inout}(a)], the gates are located on bonds starting at odd sites $i$, corresponding to pairs of sites $(i, i+1)$, whereas for odd time steps [Fig.~\ref{fig:ratio_inout}(b)], the gate positions are even. Although there are large fluctuations near the light cone boundary ($m\pm i\simeq t$ with $m$ being the site where the magic is initially inserted), the coefficients are independent of $i$ and $t$ at least well inside the light cone ($i<|m-t|$). Therefore, we further conjecture that the coefficients $\alpha(i,t)$ are independent of the position of the gate $i$ and time $t$, which is given by
\begin{eqnarray}\label{conjecture}
    \alpha(i,t)\simeq e^{-\Gamma }.
\end{eqnarray}

Substituting the above equation and Eq.~\eqref{eq:equal_output} into Eq.~\eqref{eq:single_gate_decay}, we get 
\begin{eqnarray}
   \overline{ M_i(t+1)} &=& \frac{e^{-\Gamma }}{2} \left[ \overline{ M_i(t)} + \overline{ M_{i+1} (t) } \right].
\end{eqnarray}
Then, using the relation in Eq.~\eqref{eq:totalsre_ratio}, we obtain 
\begin{eqnarray}
    \frac{\overline{ M_i(t+1)}}{\mathcal{M}(t+1)} = \frac{1}{2} \left[ \frac{\overline{ M_{i-1}(t)} + \overline{ M_{i+1} (t) }}{\mathcal{M}(t)} \right],
\end{eqnarray}
where, on the right-hand side, we use the condition $\overline{M_{i-1}(t)}=\overline{M_i(t)}$ for the output of the Clifford gate $C_{i-1, i}^{(t-1)}$ at the previous layer $t-1$ and qubit pair at sites ($i-1$, $i$) (see Fig.~\ref{fig:single_clifford}). Rewriting the above equation in terms of the normalized single-qubit averaged SRE, $a_i(t)$, we finally arrive at the discrete diffusion equation
\begin{eqnarray}
    a_{i}(t+1) = \frac{1}{2} \left[ a_{i-1}(t) + a_{i+1}(t) \right].
\end{eqnarray}

The diffusive structure of the spreading of the normalized single-qubit SRE, $a_i(t)$, is also generic for the initial product states containing a larger number of $T$ states. This is illustrated in Fig.~\ref{fig:sre_spread_tq} for an initial product state with two well-separated $T$ states, where the numerical data closely follow the discrete diffusion equation described above. The same behavior is also observed for initial states containing a few $T$ states located at consecutive sites (data not shown).

We conclude this section by contrasting the above result of the emergent diffusivity in SRE spreading with the diffusive spreading of operators observed in the systems with locally conserved quantities, such as with an $U(1)$ charge~\cite{khemani2018operator,rakovszky2018diffusive}. In these systems, a local operator that carries a conserved charge exhibits ballistic propagation of its operator front, while a substantial portion of its weight remains on operator strings that lag far behind the leading front. Because the conserved charge itself spreads diffusively, the components of the operator overlapping with the conserved sector lag behind within a region whose extent grows as $\sim \sqrt{t}$ around the origin, although the primary front advances linearly with time. This gives rise to a diffusive region near the origin, inside the overall light cone.
In contrast, we stress that our diffusive structure for the single-qubit SRE appears even though there is no such an explicit local conserved charge.

\begin{figure}[]
    \centering
    \begin{tikzpicture}
      \node[anchor=south west,inner sep=0] (img) {\includegraphics[width=0.4\textwidth,height=.21\textwidth]{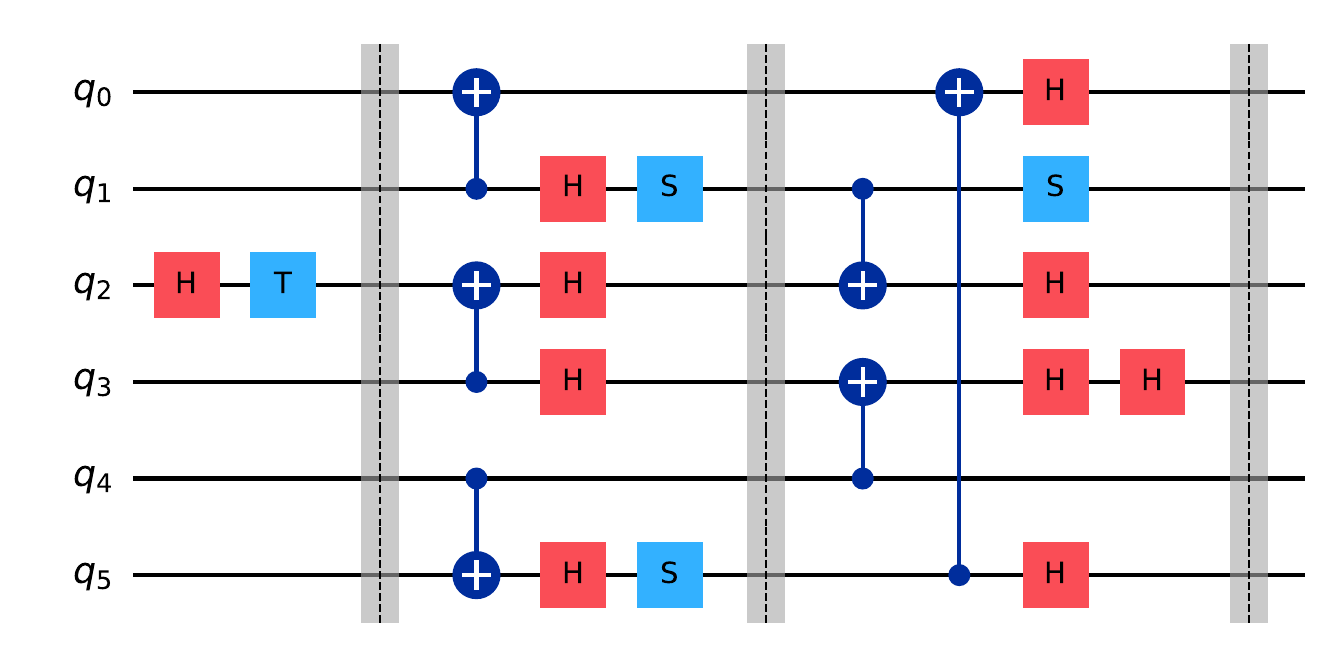}};

      \begin{scope}[x={(img.south east)},y={(img.north west)}, every node/.style={font=\large}]

        \draw[decorate,decoration={brace,amplitude=6pt}] 
          (0.3,0.95) -- (0.57,0.95)
          node[midway,above=6pt] {single depth};
        \node[below] at (0.29,0.00) {$t=0$};
        \node[below] at (0.59,0.00) {$t=1$};
        \node[below] at (0.93,0.00) {$t=2$};

      \end{scope}
    \end{tikzpicture}
    \caption{Schematic diagram for the restricted random Clifford circuit as described in Eq.~\eqref{eq:restricted_clifford}. The standard notation has been used for the two-qubit $CNOT$ gate and single-qubit gates ($H$, $S$, and $T$).  The qubit $q_2$ is prepared in a $T$-state, i.e., the initial state at $t=0$ is $\ket{00T000}$.}
    \label{fig:circuit_cnoths}
\end{figure}
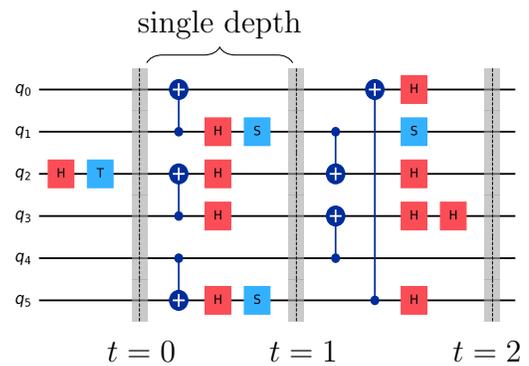

\begin{figure*}[t]
    \centering
    \includegraphics[width=.83\textwidth,height=.3\textwidth]{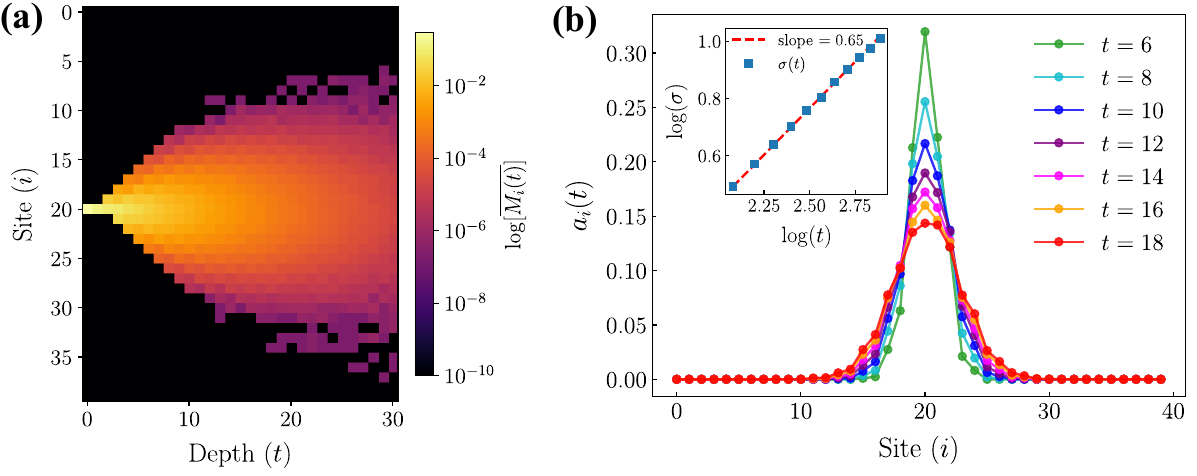}
    \caption{(a) The space-time evolution of the averaged single-qubit SRE $\overline{M_i(t)}$ and (b) the spreading profile of the normalized SRE $a_i(t)$ at different time steps $t = 6, 8, 10, 12, 14, 16$ for the restricted Clifford circuit given in Eq.~\eqref{eq:restricted_clifford}. In the inset of the panel (b), we show the scaling of the standard deviation $\sigma(t)$ of the normalized SRE $a_i(t)$ profile with time on a log-log scale. The spreading follows superdiffusive dynamics, with the standard deviation scaling as $\sigma(t) \sim t^{\beta}$, where $ \beta = 0.65$. In both panels, the results are averaged over $10^6$ realizations, and the total system size is $L=40$.}
    \label{fig:sre_restricted}
\end{figure*}

\section{Restricted Clifford circuit}\label{sec:restricted_clifford}
In this section, we investigate the robustness of the emergence of the non-ballistic spreading behavior of the normalized SRE by considering other Clifford circuits. 

Instead of considering the two-qubit Clifford gates from the whole Clifford group $\mathcal{C}_2$, we restrict ourselves to the elementary generating set of two-qubit Clifford gates consisting of the Hadamard gate ($H$), the Phase gate ($S$), and the controlled-NOT gate ($CNOT$). The circuit is constructed in a brickwork structure, with each depth step (odd or even) comprising two layers: a two-qubit layer of $CNOT$ gates followed by a single-qubit layer of $H$ and $S$ gates.

In the first layer, $CNOT$ gates are applied to the even or odd qubit pairs for odd and even depth steps, respectively. For each pair, the choice of control and target qubit for the $CNOT$ gate is randomly selected with equal probability. We denote the symmetric CNOT gate for two qubits $i$ and $j$ as $CNOT_{i \leftrightarrow j}$. 
In the subsequent layer, for each pair of qubits, two single-qubit Clifford gates are independently sampled, each chosen uniformly from $\{H,S\}$, and applied to qubits selected at random from the pair. The two draws are independent and may both target the same qubit, in which case the other qubit receives the identity.

The unitary operator $U(t', t'-1)$ in Eq.~\eqref{eq:rcbw} is now replaced as
\begin{eqnarray}\label{eq:restricted_clifford}
    U(t',t'-1) = V^{(t')}_{\mathrm{sq}} V^{(t')}_{\mathrm{CNOT}},
\end{eqnarray}
with 
\begin{eqnarray}
    V^{(t')}_{\mathrm{CNOT}} &=& \bigotimes_{\{(i,j)\}} CNOT_{i\leftrightarrow j}, \\
    V^{(t')}_{\mathrm{sq}} &=& \bigotimes_{\{(i,j)\}} \tilde{C}_{(i,j)}' \tilde{C}_{(i,j)},
\end{eqnarray}
where the qubit pairs $(i,j)$ follow the previous brickwork structure in Eq.~\eqref{eq:brickwork}, i.e.,
\begin{eqnarray}
    \{(i,j)\} &=& \{(2i, 2i+1)\} ~~~~~~~~~\text{if $t'$ is odd},\\
    \{(i,j)\} &=& \{(2i+1, 2i+2)\} ~~~~\text{if $t'$ is even},
\end{eqnarray}
for $i=0,1,\dots L-1$. Each single qubit gate $\tilde{C}_{(i,j)}$ or $\tilde{C}_{(i,j)}'$ is uniformly chosen from $\{H, S\}$ acting at a site $i$ or $j$ at random. The circuit is exemplified in Fig.~\ref{fig:circuit_cnoths}.

The numerical results for the spreading of SRE in the restricted Clifford circuit are shown in Fig.~\ref{fig:sre_restricted}. Although the SRE spreading follows the ballistic light cone at early times, it fails to reach the light cone boundaries at later times (see Fig.~\ref{fig:sre_restricted}(a)). The reason for the absence of a linear light cone will be attributed to the restriction on the choices of gates, but we leave further investigation for future problems. 

Next, we characterize the spreading behavior by looking at the spatial profile of the normalized single-qubit SRE at different time steps, as shown in Fig.~\ref{fig:sre_restricted}(b). We find that the normalized single-qubit SRE does not satisfy the discrete random equation (Eq.~\eqref{eq:dis_diff}), and it does not follow a diffusive spreading. The key reason is that the equality in Eq.~\eqref{eq:equal_output} does not hold for the restricted Clifford circuit we choose. This is because the symmetry present in the Haar random case (i.e., the group of two-qubit Clifford gates is invariant under multiplication by the two-qubit swap gate) is absent for the restricted Clifford ensemble. 

Nevertheless, by analyzing the width $\sigma(t)$ of the spreading profiles of $a_i(t)$, we find evidence of a superdiffusive structure in the spreading dynamics, as the width scales as $\sigma(t)\sim t^{\beta}$ with $\beta=0.65$. This scaling behavior is illustrated in the inset of Fig.~\ref{fig:sre_restricted}(b), where we have calculated the width of the spreading profile of $a_i(t)$ by calculating the standard deviation from the numerical data for all sites at each time step. 
Although the spatial profile of the normalized SRE does not follow the diffusive spreading, unlike the behavior observed in the Haar-random Clifford case, these results indicate that the intricate non-ballistic behavior persists in the spreading of SRE.

\begin{figure*}[t]
    \includegraphics[width=.68\textwidth,height=.5\textwidth]{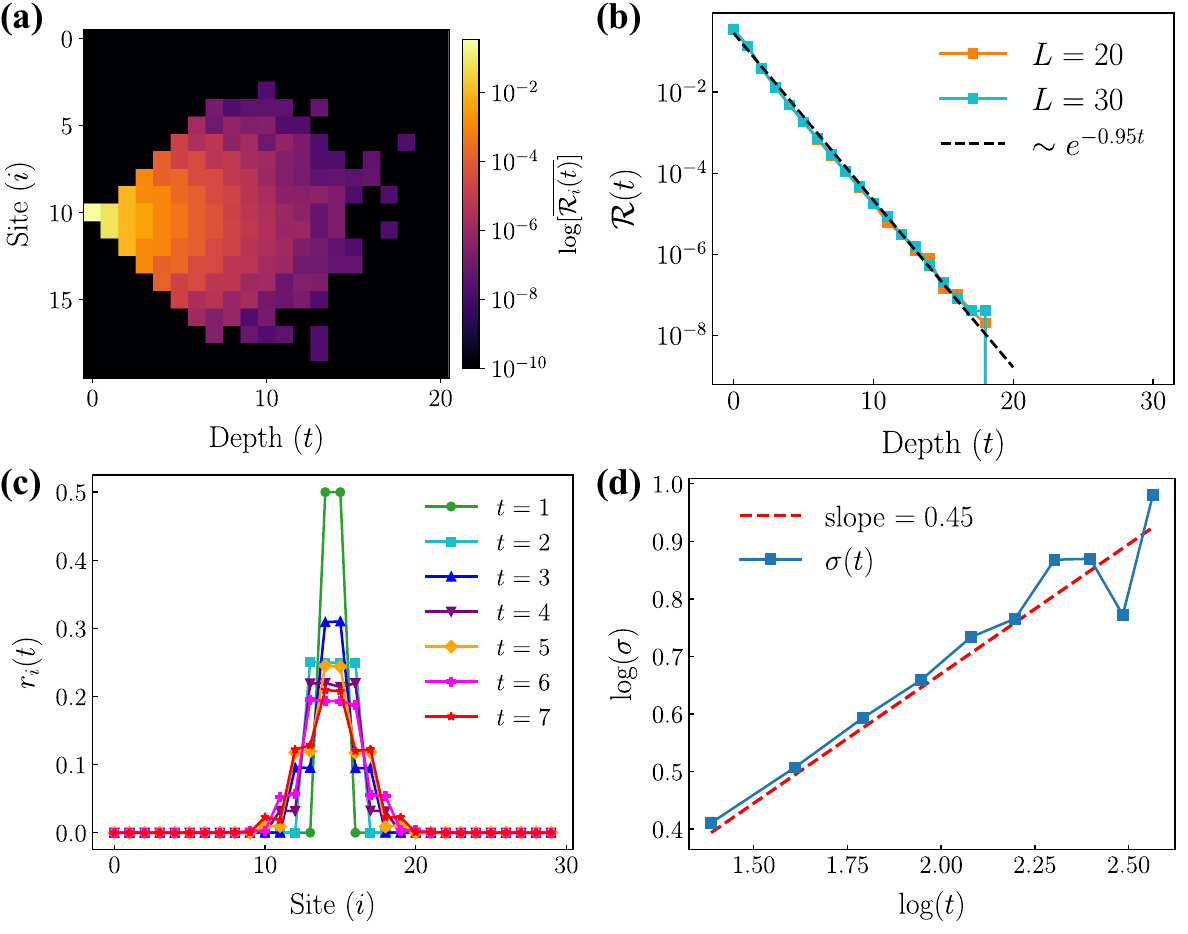}
    \caption{(a) Spatiotemporal spreading of the averaged single-qubit LROM, $\overline{\mathcal{R}_i(t)}$, for a system of size $L=20$. (b) Exponential decay of the total LROM, $\mathcal{R}(t)=\sum_i \overline{\mathcal{R}_i(t)}$, for two different system sizes $L=20$ and $L=30$. (c) Spatial profiles of the normalized single-qubit LROM, $r_i(t)=\mathcal{R}_i(t)/\mathcal{R}(t)$, at selected times $t=1,2,3,4,5,$ and $6$. (d) Time dependence of the standard deviation of the normalized LROM profiles in $\log$-$\log$ scales, demonstrating sub-ballistic (sub-diffusive) spreading. Panels (c) and (d) are shown for $L=30$. All data are averaged over $10^7$ realizations.}
    \label{fig:rom_spreading}
\end{figure*}

\section{Spreading dynamics of the robustness of magic}\label{sec:rom}

It is worth emphasizing that, although single-qubit density matrices with Pauli spectrum $\{1, 1/\sqrt{2},0,0\}$ correspond to mixed stabilizer states, the SRE assigns a nonzero value $\log(6/5)$ to this configuration, reflecting its lack of faithfulness for mixed states. To address this issue, in this section we extend our analysis to an alternative magic monotone, the robustness of magic (ROM), which is a faithful measure of magic for mixed states.

The ROM is defined as~\cite{howard2017application}
\begin{eqnarray}
    R(\rho_n) =  \min_{\{x_i\}\in\mathbb{R}^{n}}\left\{\sum_{i=1}^{n} \lvert x_i \rvert \relmiddle| \rho_n = \sum_{i=1}^{n} x_i \ket{s_i}\bra{s_i}  \right\}.
\end{eqnarray}
Here, the decompositions $\sum_i x_i \ket{s_i}\bra{s_i}$ in terms of stabilizer states $\ket{s_i} \in \mathrm{STAB}_n$ are called pseudomixtures, since the coefficients $x_i$ are not required to be nonnegative, although they satisfy $\sum_i x_i=1$. The ROM quantifies the distance
of the density matrix $\rho_n$ from the stabilizer polytope and constitutes a valid resource monotone, satisfying all axioms of the resource theory of magic for arbitrary quantum states~\cite{howard2017application}.

In our setup of local magic spreading, we focus on the ROM for single-qubit reduced density matrices. While the computation of ROM is expensive for many-qubit states, it becomes very simple for a single qubit as follows. In this case, the stabilizer polytope is given by the Bloch octahedron, which is the convex hull of the six eigenstates of the Pauli operators. Writing the single-qubit density matrix in the Pauli basis as
\begin{equation}
    \rho_i = \frac{1}{2} \sum_{\sigma_i \in \{\identity, X, Y, Z\}} c_{\sigma_i} P_{\sigma_i},
\end{equation}
the stabilizer polytope corresponds to the constraint
\begin{equation}
    \lvert c_X \rvert + \lvert c_Y \rvert + \lvert c_Z \rvert \le 1.
\end{equation}
A single-qubit state is nonstabilizer if and only if it violates this octahedral bound. Accordingly, the ROM for a single qubit can be expressed as 
\begin{equation}\label{eq:rom_qubit}
    R(\rho_i) = \max\left(1, \lvert c_{X_i} \rvert + \lvert c_{Y_i} \rvert + \lvert c_{Z_i}\rvert\right).
\end{equation}
As an example, a single qubit magic state $\ket{T}$ has the ROM value of $\sqrt{2}$.  

It is often convenient to work with the logarithm of the robustness, called as the log-free robustness of magic (LROM)~\cite{timsina2025robustness}
\begin{equation}\label{eq:lrom}
    \mathcal{R}_i(t) = \log\left[R(\rho_i(t))\right].
\end{equation}
The above quantity preserves the monotonicity properties of the measure and vanishes for all single-qubit stabilizer states, including mixed stabilizer states. Consequently, $\mathcal{R}=0$ for states associated with the Pauli spectrum $\{1, 1/\sqrt{2}, 0, 0\}$. Among the possible single-qubit Pauli spectrum listed in Table~\ref{tab:single-qubit-ps-sre}, $\mathcal{R}$ is nonzero ($\mathcal{R}=\log(\sqrt{2})$) only for the configuration $\{1, 1/\sqrt{2}, 1/\sqrt{2}, 0\}$, i.e., for the single-qubit $\ket{T}$ state.

Using the time-dependent coefficients $\{c_{\sigma_i}(t)\}$ of the single-qubit density matrix in Eq.~\eqref{eq:rom_qubit}, we compute the time-dependent single-qubit LROM $\mathcal{R}_i(t)$ for each site $i$. The numerical results of the spatiotemporal spreading of LROM are shown in Fig.~\ref{fig:rom_spreading}. The spreading of the averaged single-qubit LROM $\mathcal{R}_i(t)$ exhibits a clear ballistic light cone as illustrated in Fig.~\ref{fig:rom_spreading}(a), and the sum of all single-qubit LROM $\mathcal{R}(t)=\sum_i \mathcal{R}_{i}(t)$ decays exponentially with time (see Fig.~\ref{fig:rom_spreading}(b)), the behavior similar to that observed for the SRE. The spreading profiles of the normalized single-qubit LROM are shown in Fig.~\ref{fig:rom_spreading}(c). Unlike the SRE, the spreading profiles for the normalized LROM, $r_i(t)=\mathcal{R}_i(t)/\mathcal{R}(t)$, exhibit nonmonotonic peaks: the maximum value of the profile decreases in a nonmonotonic manner with time and therefore does not obey a discrete diffusion equation. Nevertheless, the spreading remains non-ballistic (sub-diffusive). Specifically, the standard deviation of the LROM profiles scales as $\sigma(t) \sim t^{\beta}$ with $\beta \approx 0.45$, as shown in Fig.~\ref{fig:rom_spreading}(d), indicating slower-than-diffusive dynamics.

\section{Operational implications}\label{sec:implications}
We next provide an operational implication of the single-qubit SRE based on the setup we use, especially for the case where a single-qubit $\ket{T}$ state is inserted in the initial state and it evolves under the random Clifford circuit.

It is important to note that for a particular realization of the random circuit, the nonzero single-qubit magic for the site $i$ refers to a global pure state having the following factorized form:
\begin{equation}
    \ket{\psi} = \ket{T}_i \otimes \ket{\phi}_{\bar{i}},
\end{equation}
where the magic state $\ket{T}_i$ is at the $i$-th site and $\ket{\phi}_{\bar{i}}$ is a stabilizer state for the complement part of the qubit $i$. In other words, the magic for a single-qubit reduced density matrix will be nonzero only if there is a pure magic state $\ket{T}$ that is in a direct product state with the stabilizer state for the rest of the system. This is clearly evident from the allowed single-qubit Pauli spectrum listed in Table~\ref{tab:single-qubit-ps-sre}, as there is only one configuration that has nonzero magic, which is nothing but a $\ket{T}$ state. We note that the above direct-product structure of a single-qubit pure magic state is sometimes a primary goal in a magic-state distillation process and is also important for implementing a quantum task in quantum devices.

Now, let us define $p_i(t)$ as the probability of finding the $\ket{T}_i$ state at the $i$-th site at time $t$ with respect to random Clifford circuits. We also define $q_i(t)$ as the probability corresponding to the case where we find a mixed stabilizer state with a Pauli spectrum $\left\{1,1/\sqrt2,0,0\right\}$. Then, the averaged single-qubit SRE is given by
\begin{equation}
    \overline{M_i\left(t\right)} = p_i\left(t\right)\log{(4/3)}+q_i\left(t\right)\log{(6/5)}.
\end{equation}
Therefore, the averaged single-qubit SRE operationally tells the upper bound on the probability of finding a pure magic $\ket{T}$ state at site $i$ under the random dynamics of a Clifford circuit, i.e.,
\begin{equation}
    p_i\left(t\right)\le\frac{\overline{M_i\left(t\right)}}{\log{(4/3)}}.
\end{equation}
Since our numerical results for the Haar-random Clifford circuits suggest that the averaged single-qubit SRE has the following form,
\begin{eqnarray}
    \overline{M_i\left(t\right)}\simeq\frac{e^{-\Gamma t}}{\sqrt{2\pi t}}e^{-\frac{\left|i-m\right|^2}{2t}},
\end{eqnarray}
where $m$ is the qubit where the magic state is inserted in the initial state, we have
\begin{eqnarray}
    p_i\left(t\right)\lesssim\frac{1}{(\log{4/3)}}\frac{e^{-\Gamma t}}{\sqrt{2\pi t}}e^{-\frac{\left|i-m\right|^2}{2t}}.
\end{eqnarray}
This indicates that the probability of finding the pure magic state is exponentially suppressed in time and that it is further suppressed for qubits that are distant from the site $m$ with the Gaussian dependence. 
Moreover, a similar analysis of LROM leads to the fact that the exact probability of finding the single-qubit magic state is given by $p_i(t)=\overline{\mathcal{R}_i(t)}/\log{(\sqrt{2})}$. This  also decays exponentially in time and additionally shows the non-ballistic structure inside the light cone as shown in Fig.~\ref{fig:rom_spreading}.

\section{Conclusions}
\label{sec:discussion}
In this study, we reveal a rich dynamical structure underlying the spreading of stabilizer R\'enyi entropy (SRE) in many-body systems within a minimal setup of Clifford circuits. By injecting a few local non-stabilizer states into an otherwise stabilizer product state, we study the propagation of SRE under a brickwork random Clifford circuit. While the unitary Clifford evolution preserves the total SRE of the pure global state, we showed that the local structure of single-qubit SRE evolves in a nontrivial way as the correlations spatially spread across the system.

We observe that the sum of the single-qubit averaged SRE decays exponentially in time. This decay indicates that as the state becomes more entangled, the magic is increasingly encoded in nonlocal entanglement. However, we also discover an interesting spatial structure in the spreading dynamics of the single-qubit SRE.
This becomes evident by introducing a normalized profile of the single-qubit averaged SRE; we indeed reveal an underlying diffusive structure in SRE spreading within the light cone for a generic local random Clifford circuit.
We find that the diffusive equation holds true with high accuracy, which we have explained through the analysis of a single two-qubit Clifford gate from the circuit with the conjecture in Eq.~\eqref{conjecture}.
Additionally, we investigate the robustness of the non-ballistic structure of SRE spreading in relation to other Clifford circuits by analyzing a restricted Clifford circuit where the gates are not drawn from the full Clifford group. We unveil that the spreading structure becomes superdiffusive in this restricted class of Clifford circuits.

We further extend our analysis of the nonballistic spreading of another magic monotone, i.e., the log-free robustness of magic (LROM), which is a faithful measure applicable to mixed states. We find that the LROM exhibits similarly intricate non-ballistic dynamics within the interior of the light cone, closely paralleling the behavior observed for the SRE. Therefore, the observed dynamical features in local spreading of magic are robust across distinct measures of magic.

The resulting spreading dynamics of the SRE admit a clear operational meaning in terms of extractability of single-qubit magic state under the dynamics of Clifford circuit. In particular, the averaged single-qubit SRE, $\overline{M_i(t)}$ provides an upper bound on the the probability of finding a pure single-qubit magic state on qubit $i$, corresponding to a factorized global pure state of the form $\ket{T}_i \otimes \ket{\phi}_{\bar{i}}$. The exponential decay of $\overline{M_i(t)}$ indicates that the probability of finding a pure magic state decays rapidly with time as magic becomes delocalized into nonlocal correlation under the scrambling dynamics of the Clifford circuit. In addition to this exponential decay, the non-ballistic spatial structure observed reveals a higher probability in the vicinity of the initially injected qubit than that for the distant qubit.

Our work is relevant on how local magic spatially spreads under a magic-conserving dynamics and fundamentally important to understand the thermalization dynamics of many-body systems, which is in principle tested on a quantum computer.
Moreover, it suggests several promising directions for future research. First, a more profound quantitative and analytical understanding of the observed diffusive spreading (including the conjecture in Eq.~\eqref{conjecture}) is scope for future research. Second, unveiling the universality class in the normalized SRE dynamics, whether diffusive, superdiffusive, subdiffusive, or ballistic, is also an important direction of this research. Another compelling avenue is to understand how the spreading behavior of the SRE is altered in the presence of measurement and dissipation.

\begin{acknowledgments}
We thank Hishanori Oshima and Yuuya Chiba for discussions and valuable comments during the project. S.M. acknowledges Alioscia Hamma, Stefano Cusumano, and Daniele Iannotti for helpful comments.  The numerical calculations in this work were carried out using Qiskit libraries~\cite{qiskit2024}.
This work was supported by JST ERATO Grant Number JPMJER2302, Japan. R.H. was supported by JSPS KAKENHI Grant No. JP24K16982.\\
\end{acknowledgments}

\textit{Note Added:} After we submitted the first version of our manuscript, two independent studies by Bejan, Claeys, Yao~\cite{bejan2025magicspreading} and Aditya, Turkeshi, Sierant~\cite{aditya2025resourcesspreading} appeared, which investigate local magic spreading under Clifford gates. While they use different approaches from ours, the results are consistent with one another where they overlap.

\appendix
\section{SRE for a single-qubit magic state} \label{app:magic_tstat}
Here, we calculate the SRE for a single-qubit magic state
\begin{eqnarray}
    \ket{T}=\frac{1}{\sqrt{2}}\left(\ket{0}+e^{i\pi/4}\ket{1}\right).
\end{eqnarray}
The expectation values of the Pauli operators on the above state are given by
\begin{eqnarray}
    \bra{T}\identity\ket{T}&=&1,\\
    \bra{T}X\ket{T}&=&\frac{1}{\sqrt{2}},\\
    \bra{T}Y\ket{T}&=&-\frac{1}{\sqrt{2}},\\
    \bra{T}Z\ket{T}&=&0.
\end{eqnarray}
Consequently, the Pauli spectrum of the magic state $\ket{T}$ is given by $\{1, 1/\sqrt{2}, 1/\sqrt{2}, 0\}$, and in the Pauli basis, the state $\ket{T}$ can be expressed as
\begin{equation}
    \ket{T} \bra{T} = \frac{1}{2} \left[ \identity + \frac{1}{\sqrt{2}} \left(X-Y\right) \right].
\end{equation}
Using Eq.~\eqref{eq:sre_alpha} in the main text, one can easily obtain the $\alpha$-SRE for the magic state is given by
\begin{eqnarray}
    M^{(\alpha)}(\ket{T})&=&\frac{1}{1-\alpha}\log \left[\frac{1}{2}\left(1+\frac{2}{2^\alpha}\right)\right], \\
\end{eqnarray}
For the R\'enyi index $\alpha=2$, we have
\begin{eqnarray}
    M(\ket{T})&=&\log\left(\frac{4}{3}\right).
\end{eqnarray}

\section{Elements of a \texorpdfstring{$L$}{L}-qubit Pauli spectrum}\label{app:ps}
We now extend the single-qubit spectrum calculation to a two-qubit case. Consider the initial state $\ket{\psi_0}=\ket{0T}$. The expectation values for 
the two-qubit Pauli strings  $P_{\sigma_i \sigma_{i+1}}=\left\{\identity\identity, \identity X, \identity Y,\dots ZZ\right\}$ with respect to the state $\ket{\psi_0}$ have the following nonzero components:
\begin{eqnarray}\label{eq:two-qubit-ps-1}
    \bra{0T} \identity\identity\ket{0T}&=& \bra{0T} Z \identity \ket{0T} =1, \\\label{eq:two-qubit-ps-2}
        \bra{0T} \identity X \ket{0T}&=& \bra{0T} ZX \ket{0T} = \frac{1}{\sqrt{2}}, \\\label{eq:two-qubit-ps-3}
    \bra{0T} \identity Y \ket{0T}&=& \bra{0T} ZY\ket{0T} = - \frac{1}{\sqrt{2}},
\end{eqnarray}
while the expectation values of the remaining ten Pauli strings vanish.

This analysis can easily be generalized to an $L$-qubit system. For the initial state $\ket{\psi_0}=\ket{000\cdots 0T0\cdots 0}$, where a $T$ state is located at site $m$, the expectation values of the $L$-qubit Pauli strings are given by,
\begin{align}
    \bra{\psi_0} A_{0}\cdots A_{m-1}\identity_m A_{m+1}\cdots A_{L-1} \ket{\psi_0} &= 1, \label{eq:ps_expectation}\\
    \bra{\psi_0} A_{0}\cdots A_{m-1}X_m A_{m+1}\cdots A_{L-1} \ket{\psi_0} &= \frac{1}{\sqrt{2}}, \label{eq:ps_expectation2}
    \\
    \bra{\psi_0} A_{0}\cdots A_{m-1}Y_m A_{m+1}\cdots A_{L-1} \ket{\psi_0} &= -\frac{1}{\sqrt{2}}, \label{eq:ps_expectation3}\\
    \bra{\psi_0} A_{0}\cdots A_{m-1}Z_m A_{m+1}\cdots A_{L-1} \ket{\psi_0} &= 0, \label{eq:ps_expectation4}
\end{align}
where $A_i = \{\identity, Z\}$. This result shows that for an initial product state containing a single $T$-state excitation, the elements of a Pauli spectrum can only take three distinct values: $0$, $1$, and $1/\sqrt{2}$.

\section{Possible Pauli spectra for a physically allowed density matrix}\label{app:allowed_ps}
In this appendix, we analyze the possible Pauli spectra that emerge at the output of an arbitrary two-qubit Clifford gate $C_{i,i+1}^{(t)}$ within the brickwork circuit.
We consider the two-qubit density matrix $\rho_{i,i+1}$ defined on the Hilbert space of sites $i$ and $i+1$. The Pauli decomposition of  $\rho_{i,i+1}$ is given by
\begin{eqnarray}\label{eq:tq-den-mat}
    \rho_{i,i+1}(t) = \frac{1}{4} \sum_{\sigma_i, \sigma_{i+1}} c_{\sigma_i \sigma_{i+1}} (t) P_{\sigma_i, \sigma_{i+1}}
\end{eqnarray}
where $P_{\sigma_i, \sigma_{i+1}}$ are two-qubit Pauli strings, and the coefficients $c_{\sigma_i, \sigma_{i+1}} (t)$ are given by
\begin{eqnarray}
    c_{\sigma_i, \sigma_{i+1}} (t)= \bra{\psi_0}P_{\sigma_i, \sigma_{i+1}}(t)\ket{\psi_0}.
\end{eqnarray}
These coefficients constitute the time-dependent Pauli spectrum, $\{\lvert c_{\sigma_i, \sigma_{i+1}} (t) \rvert\}$, for the two-qubit reduced density matrix $\rho_{i,i+1}(t)$. From Eq.~\eqref{eq:tq-den-mat}, the reduced single-qubit density matrices can be obtained by tracing out the other qubit:
\begin{eqnarray}
    \rho_i(t) &=& \frac{1}{2} \sum_{\sigma_i \in \{\identity, X,Y,Z\} } c_{\sigma_i \identity}(t) \sigma_i, \\
    \rho_{i+1}(t) &=& \frac{1}{2} \sum_{\sigma_{i+1} \in \{\identity, X,Y,Z\} } c_{\identity\sigma_{i+1}}(t) \sigma_{i+1}.
\end{eqnarray}

To ensure that the operators $\rho_{i, i+1}(t)$, $\rho_{i}(t)$, and $\rho_{i+1}(t)$ are  valid density matrices, certain constraints are imposed on $c_{\sigma_i\sigma_{i+1}}(t)$.
For example, they are real and bounded as  $|c_{\sigma_i\sigma_{i+1}}(t)|\le 1$. The normalization condition, $\Tr[\rho_{i,i+1}]=\Tr[\rho_{i}]=\Tr[\rho_{i+1}]=1$, is fixed by $c_{\identity\identity}(t)=1$. Considering the purity of the reduced density matrices, we also have the following conditions:
\begin{eqnarray}
    \sum_{\sigma_i,\sigma_{i+1}} \lvert c_{\sigma_i \sigma_{i+1}}(t)\rvert^2 &\leq& 4, \\
    \sum_{\sigma_i} \lvert c_{\sigma_i \identity}(t) \rvert^2 &\leq& 1, \\
    \sum_{\sigma_{i+1}} \lvert c_{\identity \sigma_{i+1}}(t)\rvert^2 &\leq& 1.
\end{eqnarray}
Imposing these conditions on the sixteen coefficients $c_{\sigma_i \sigma_{i+1}}(t)$, restricting their values to the discrete set $\{\pm1, \pm1/\sqrt{2},0\}$ (see Appendix~\ref{app:ps}), and enforcing the positive semidefinite condition for valid density matrices, we numerically obtained all possible Pauli spectra consistent with the initial state and the brickwork Clifford circuit. 

\begin{table}
\centering
\caption{Distinct combinations of $(a,b)$, where $a$ and $b$ denote the number of elements with values $1$ and $1/\sqrt{2}$, respectively, in the two-qubit Pauli spectrum $\{\lvert c_{\sigma_i, \sigma_{i+1}}(t) \rvert\}$, together with the corresponding two-qubit SRE values $M_{i,i+1}(t)$.}
\begin{ruledtabular}
\begin{tabular}{cc}
\textrm{$(a,b)$} & \textrm{$M_{i,i+1}(t)$} \\[2pt]
\colrule
(1, 0) & $0$ \\
(2, 0) & $0$ \\
(4, 0) & $0$ \\
(1, 1) & $\log(6/5)$ \\
(1, 2) & $\log(4/3)$ \\
(2, 2) & $\log(6/5)$ \\
(2, 4) & $\log(4/3)$ \\
\end{tabular}
\end{ruledtabular}
\label{tab:twoqubit_sre}
\end{table}

\begin{table}
\centering
\caption{Possible single-qubit Pauli spectra and the corresponding SRE values $M_i(t)$.}
\begin{ruledtabular}
\begin{tabular}{lc}
\textrm{Single-qubit Pauli spectrum} & \textrm{$M_i(t)$} \\[2pt]
\colrule
$\{1,0,0,0\}$               & $0$ \\
$\{1,1,0,0\}$               & $0$ \\
$\{1,1/\sqrt{2},0,0\}$             & $\log(6/5)$ \\
$\{1,1/\sqrt{2},1/\sqrt{2},0\}$           & $\log(4/3)$ \\
\end{tabular}
\end{ruledtabular}
\label{tab:single-qubit-ps-sre}
\end{table}

Analogous to the single-qubit SRE in  Eq.~\eqref{eq:sre_coeff}, we define the two-qubit SRE, denoted by $M_{i,i+1}(t)$, based on the two-qubit Pauli spectrum $\{\lvert c_{\sigma_i, \sigma_{i+1}} (t) \rvert\}$. Denoting the number of elements in the Pauli spectrum with values $1$ and $ 1/\sqrt{2}$, as $a$ and $b$, respectively, the possible distinct two-qubit Pauli spectrum and the corresponding two-qubit SREs are summarized in Table~\ref{tab:twoqubit_sre}. For completeness, the possible single-qubit Pauli spectrum and corresponding single-qubit SRE values are listed in Table~\ref{tab:single-qubit-ps-sre}.

\bibliography{references_sre.bib}
\end{document}